\documentclass[12pt,a4paper,final]{iopart}

\usepackage{iopams}  
\usepackage{graphicx}
\usepackage[breaklinks=true,colorlinks=true,linkcolor=blue,urlcolor=blue,citecolor=blue]{hyperref}
\usepackage{ulem}

\newcommand{\be}{\begin{eqnarray}}
\newcommand{\ee}{\end{eqnarray}}

\renewcommand{\theequation}{\arabic{equation}}

\begin{document}

\title[Localization and slow-thermalization in a cluster spin model]
{Localization and slow-thermalization in a cluster spin model}

\author{Yoshihito Kuno$^{1}$, Takahiro Orito$^{2}$ and Ikuo Ichinose$^{3}$}
\address{$^1$ Graduate School of Engineering Science, Akita University, Akita 010-8502, Japan}
\address{$^2$ Graduate School of Advanced Science and Engineering, Hiroshima
 University, 739-8530, Japan}
\address{$^3$ Department of Applied Physics, Nagoya Institute of Technology, Nagoya, 466-8555, Japan}
%\address{Address One, Japan}
%\address{$^2$Address Two, Neverland}
\ead{kuno421yk@gmail.com}

\begin{abstract}
Novel cluster spin model with interactions and disorder is introduced and studied. 
In specific type of interactions, 
we find an extensive number of local integrals of motion (LIOMs), which are a modified version of the stabilizers in quantum information,
i.e., mutually commuting operators specifying all quantum states in the system. 
These LIOMs can be defined for any strength of the interactions and disorder,
and are of compact-support instead of exponentially-decaying tail.
Hence, even under the presence of interactions, integrability is held, and
all energy eigenstates are labeled by these LIOMs and can be explicitly obtained.
Integrable dynamics is, then, expected to occur. 
The compact-support nature of the LIOMs crucially prevents the thermalization and entanglement spreading. 
We numerically investigate dynamics of the system governed by the existence of
the compact-support LIOMs, and clarify the effects of additional interactions, 
which break the compact-support nature of the LIOMs. 
There, we find that the ordinary MBL behaviors emerge, such as the logarithmic growth
of the entanglement entropy in the time evolution.
Besides the ergodicity breaking dynamic, we find that symmetry-protected-topological 
order preserves for specific states even in the presence of the interactions.
\end{abstract}

%Uncomment for PACS numbers title message
%\pacs{00.00, 20.00, 42.10}
%\pacs{67.85.-d, 03.75.Lm, 05.30.Jp, 73.21.Cd}
% Keywords required only for MST, PB, PMB, PM, JOA, JOB? 
\vspace{2pc}
\noindent{\it Keywords}: Many-body localization, cluster spin model, thermalization.
% Uncomment for Submitted to journal title message
%\submitto{\NJP}
% Comment out if separate title page not required
\maketitle

%%%%%%%%%%%%%%%%%%%%%%%%%%%%%%%%%%%%%%%%%%%%%%%%%%%%%%%%%%%%%%%%
\section{Introduction}\label{intro}
Localization and thermalization are deeply related notions~\cite{Nandkishore2015,Abanin}. 
In a closed system separated from environment, if all states are localized, 
the system does not thermalize. This is observed in quench dynamics. 
This character persists even in interacting many-body systems. This phenomenon is called many-body-localization 
(MBL) \cite{Basko2006,Bardarson}. 
Such non- or slow-thermalization dynamics is a novel example for the breaking of 
the eigenstate thermalization hypothesis \cite{Nandkishore2015,Rigol2007}. 
%\out{Also, weak localization and weak ergodicity breaking in some specific models such as quantum scar states 
%are now getting a lot of attention
%\cite{Bernien,Turner2018,Choi2019,Ho2019,Iadecola2020,Bluvstein2020,McClarty2020,Shibata2020,Lee2020,KMH2020,Serbyn2021,Jeyaretnam,KMH2020,KMH2021}}.
What types of models exhibit localization or MBL phenomena, and what constraints induce them are important questions. A key concept for the questions is an extensive number of local integrals of motions (LIOMs)
\cite{Serbyn2013,Huse2014,Imbrie2016,Imbrie2017}.
Investigating the LIOMs in specific models is important and useful to scrutinize localization phenomena
from the general perspective.
Constructing some classification scheme of localization from the bottom-up approach
by the study on various concrete examples is an important subject \cite{Serbyn2021}. 

In this article, we study an extended version of the cluster spin model, 
in which LIOMs are obtained explicitly. The original cluster model \cite{Briegel2001} plays an important role in the context of 
quantum computation and topological study in condensed 
matter physics \cite{Pachos2004,Son,Smacchia}. 
This model exhibits localization phenomena with symmetry-protected-topological (SPT) order, as well as ergodicity breaking dynamics. 
Even in the presence of interactions, which preserve the symmetries of the system, 
localization persists, namely, topological MBL takes place there \cite{Bauer2013,Bahri,Vasseur2016,Parameswaran,Decker,Kuno2019,Wahl2020_1,Wahl2020_2,Wahl2020_3,Kemp2020,Sahay2021,Duque2021}.

The LIOMs in the cluster-spin model are of compact-support.
There are extensive works on various systems with compact-support
LIOMs including the Creutz ladder, diamond lattice, etc \cite{KOI2020,Danieli_1,Roy,Zurita,OKI2021,Tilleke,Khare}.
Important insight into localization has been obtained by the studies on these systems.
The present study also belongs to the category of these works. Furthermore, important relationship between many-body physics
and quantum information theory attracts lots of attentions nowadays. 
In this context, compact-support LIOMs are nothing but generators of a stabilizer  
group, which play an important role in quantum error-correcting code
\cite{Huse2020,Fisher2021,Lavasani2021}.
From this point of view, study on localization with compact-support LIOMs is important
and useful.
In fact as we show later on, modified LIOMs obtained in this work may define a new type of 
stabilizer group.

The extended cluster spin model, which we propose,
includes a certain type of interactions and disorder, and
possesses an extensive number of modified LIOMs. 
The modified LIOMs have compact-support, which are different from Anderson orbital defined on disordered systems or the LIOMs ($\ell$-bit)
in conventional MBL, both of which have exponentially-decaying support \cite{Nandkishore2015}.
The compact-support of the LIOMs is related to the fact that the interactions only 
locally mix many-body states. 
The modified LIOMs are also regarded as an extended counterpart of 
the stabilizers in the original cluster spin model. 
They label all eigenstates of the interacting and disordered Hamiltonian. 
Then, we show that the presence of the modified LIOMs induces specific localized phenomena originated from the integrability by the compact-support LIOMs, i.e.,
slow-thermalization (ergodicity breaking dynamics). There are many quantities to evaluate the slow-thermalization and ergodicity breaking dynamics, 
such as fluctuations of local observables, the inverse participation ratio,
Kullback-Leibler divergence, mutual information, overlap correlation function, etc
\cite{Luitz2015,Evers2008}. 
In this work, we mainly investigate return probability and
entanglement entropy (EE). 
If  system exhibits a slow-thermalization and ergodicity breaking
dynamics, EE does not increase for a specific initial state and return probability
remains a finite value. 
These properties are numerically demonstrated in this article.
We investigated how the dynamics of the system changes under additional Ising-type interactions, with which the compact-support LIOMs cannot be clearly defined.
The specific localization behavior may be destroyed by such interactions. 
We numerically investigate these problems. 
In addition, we qualitatively investigate the stability of the SPT order, which is
characterized by a string order in the original cluster model \cite{Smacchia}, 
by introducing an extended string order. 
We observe the SPT order tends to be stable in the extended cluster spin model with or without the Ising-type interactions.

The rest of the paper is organized as follows. 
In Sec.~\ref{model}, we introduce the cluster spin model and its extended version and 
comment on its basic properties. 
In Sec.~\ref{SecIII}, we introduce the modified LIOMs defined for the extended cluster spin model. 
The LIOMs are exactly and explicitly obtained even in the presence of disorder. 
We further give a Majorana-fermion representation of the model to understand the target model
clearly. 
In Sec.~IV, we show the study of a small system and numerically demonstrate the presence of 
the modified LIOMs in detail.
In Sec.~V, we present numerical observation of characteristic dynamics originated from the integrability by the modified LIOMs. 
In Sec.~VI we move on to the case with the additional 
Ising-type interactions, where the modified LIOMs are no longer to be defined. 
There, we numerically observe the ordinary MBL phenomena.
Section VII is devoted to discussion and conclusion.

%%%%%%%%%%%%%%%%%%%%%%%%%%%%%%%%%%%%%%%%%%%%%%%%%%%%%%%%%%%%
\section{Model}\label{model}
We focus on an extended version of the cluster spin model with interactions and disorder. 
The cluster spin model is a basic model to implement a measurement based 
quantum computation in quantum information theory and 
also exhibits a SPT phase.
These properties were extensively studied in \cite{Briegel2001,Pachos2004,Son,Smacchia}.
In this article, we consider the following extended version of the cluster spin model 
defined on the one-dimensional lattice,
\begin{eqnarray}
&&H=\sum^{L/2-1}_{\ell=0}[J_{2\ell}K_{2\ell}+J_{2\ell+1}K_{2\ell+1}]+H_{\rm int},\\
&&H_{\rm int}=\sum^{L/2-1}_{\ell=0}gV_{\ell},
\label{Model0}
\end{eqnarray}
where $K_{j}=\sigma^{z}_{j-1}\sigma^{x}_j\sigma^{z}_{j+1}$ is a stabilizer operator 
composed of Pauli matrices, 
$V_{\ell}=N_{\ell-1}[K^{+}_{2\ell}K^{-}_{2\ell+1}+K^{-}_{2\ell}K^{+}_{2\ell+1}]$, $N_{\ell}=K_{2\ell}+K_{2\ell+1}$ and 
$K^{\pm}_{j}=\frac{1}{2}(\sigma^{z}_{j}\mp i\sigma^{z}_{j-1}\sigma^{y}_{j}\sigma^{z}_{j+1})$.
$K^{\pm}_{j}$ is a raising and lowering operator for cluster state \cite{Jeyaretnam}, 
$J_j$ is site-dependent scalar potential of the stabilizer $K_j$, 
and $g$ is strength of the interactions. Note that the complicated interaction $H_{\rm int}$ has possibility to be implemented in some quantum circuits in the context of quantum simulation by quantum computer \cite{Smith2020}. 
The stabilizer $K_j$ is a dressed spin, which satisfies $K^{2}_j=1$ and $SU(2)$-spin algebra,
\begin{eqnarray}
[K_j,K^{\pm}_{j}]=\pm2 K^\pm_j.
\label{Model0}
\end{eqnarray}
The Hamiltonian $H$ has $\mathbb{Z}_{2}\times \mathbb{Z}^{T}_{2}$ symmetry, which is composed of 
the global spin flip $\prod^{L-1}_{j=0}\sigma^{x}_j$ and the complex conjugation 
(time-reversal operation) \cite{Verresen2017,Verresen2018,Smith2020}. 

For $g=0$, the Hamiltonian $H$ is the original cluster spin model, where 
the set of the stabilizers $\{K_j\}$ become an extensive number of LIOMs since 
$[H_{g=0},K_j]=0$ and $[K_j,K_{j'}]=0$ for any $j$ and $j'$. If the site-dependent couplings $\{ J_j\}$ are negative, the unique ground state appears,
where the state is labeled by $K_j=1$. 
This unique ground state is called a cluster state having short-range entanglement and is regarded 
as a SPT state \cite{Son,Smacchia,Verresen2017,Verresen2018,Smith2020}. 
The cluster state is exactly written by 
\begin{eqnarray}
|\Psi_{cl}\rangle=\sqrt{2^{L}}\biggl[\prod^{L-1}_{j=0}K^{+}_j\biggr]|\Uparrow\rangle,\:\:
|\Uparrow\rangle =\bigotimes^{L-1}_{j=0}|\uparrow \rangle_j,
\label{exact}
\end{eqnarray}
where $\sigma^{z}_j|\uparrow\rangle_j=|\uparrow\rangle_{j}$, and 
$|\Uparrow\rangle$ is a ferromagnetic state with all spins up.
The topological property of this state is known to be robust against perturbations preserving 
the symmetries such as $\mathbb{Z}_{2}\times \mathbb{Z}_{2}$ \cite{Smacchia} and $\mathbb{Z}_{2}\times \mathbb{Z}^{T}_{2}$ \cite{Smith2020}, and the short-range entanglement character is preserved unless energy gap closes and 
a phase transition to a symmetry-breaking phase takes place. 

It is expected that the quantum order of ground state such as long-range order or topological properties as in Eq.(\ref{exact}) can survive even in the excited states under disorders \cite{Huse2013}. In our model, when the site-dependent couplings $\{ J_{j}\}$ ($j=0,1,\cdots, L-1$) are randomly varied including sign change, 
the system exhibits certain localization properties
with short-range entanglement of the cluster state, where every eigenstate is 
uniquely labeled by the set of eigenvalues of the LIOMs, $K_j=\pm 1$. 
The phenomenon is stable against perturbations 
preserving the symmetry (such as Ising interactions), 
namely, the topological MBL, which has been studied extensively \cite{Bahri,Decker,Vasseur2016,Parameswaran,Kuno2019,Wahl2020_1,Wahl2020_2,Wahl2020_3,Kemp2020,Sahay2021,Duque2021}.

In this article, we focus on the case of $g\neq 0$ and also a specific
type of disorder such as $J_{2\ell}=-J_{2\ell+1}=\lambda_{\ell}$, 
where $\lambda_{\ell}$'s are uniformly-distributed random variables, $\lambda_{\ell} \in [-W,W]$.
In this case, the system has modified LIOMs as we show in the following section.

\section{Modified local integrals of motion}\label{SecIII}
\begin{figure}[h]
\begin{center} 
\includegraphics[width=10cm]{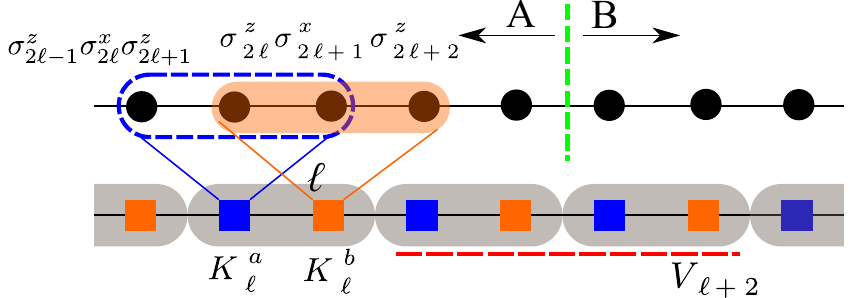}  
\end{center} 
\caption{Schematic figure of the model of Eq.~(\ref{Model_mod}). 
The upper chain is the original spin-lattice. The original spin-lattice is mapped onto the cluster-basis lattice
in the lower chain. The red dashed horizontal line represents the regime where each term
of the interaction $V_{\ell}$ acts. 
The green vertical dashed line represents 
the entanglement cut for a pair of subsystems in the calculation of the EE.}
\label{Fig0}
\end{figure}

In this section, we show the existence of a novel type of LIOMs 
for finite $g$.
Under the disorder $\{ \lambda_{\ell}\}$, the target cluster model is given as
\begin{eqnarray}
H=H_0+H_{\rm int},\:\:H_0=\sum^{L/2-1}_{\ell=0}\lambda_{\ell}[K^{a}_{\ell}-K^{b}_{\ell}],
\label{Model_mod}
\end{eqnarray}
where we have introduced unit-cell including two sites $2\ell$ and $2\ell+1$, and 
``$a$" and ``$b$" indices for even and odd sites in the unit-cell. 
The operators $K_{2\ell (2\ell+1)}$ are relabeled as $K^{a(b)}_{\ell}$. 
The schematic lattice structure is shown in Fig.~\ref{Fig0}. 
Similar structure to $K^{a(b)}_{\ell}$ was used in other lattice models, e.g., 
for compact localized states \cite{KOI2020,OKI2020,OKI2021}.

%%%%%%%%%%%%%%%%%%%%%%%%%%%%%%%%%%%% 
%%%%%%%%%%%%%%%%%%%%%%%%%%%%%%%%%%%%
By following our previous study on the Creutz ladder \cite{OKI2020}, 
we find that there are operators commuting with the total Hamiltonian $H$, 
which are given by 
\begin{eqnarray}
\tilde{K}^a_{\ell}=K^{a}_{\ell}+\frac{g}{2\lambda_{\ell}}V_{\ell},\:\:
\tilde{K}^b_{\ell}=K^{b}_{\ell}-\frac{g}{2\lambda_{\ell}}V_{\ell},
\label{LIOM2}
\end{eqnarray}
and the total Hamiltonian is expressed as
$H=\sum^{L/2-1}_{\ell=0}\lambda_{\ell}[\tilde{K}^{a}_{\ell}-\tilde{K}^{b}_{\ell}]$. 
These operators are of compact-support and satisfy $[\tilde{K}^{a(b)}_{\ell},\tilde{K}^{a(b)}_{\ell '}]=0$ for any $\ell$ and $\ell '$, and
$[H,\tilde{K}^{a(b)}_{\ell}]=0$. 
That is, these operators are nothing but LIOMs, which are a modified version of the original stabilizers $K_{j}$. 
Also, note that the LIOMs, $\tilde{K}^{a(b)}_{\ell}$, can be defined for any disorder realizations and interaction 
strength, $\{\lambda_\ell\}$ and $g$, 
whereas $(\tilde{K}^{a(b)}_{\ell})^{2}\neq 1$, that is, 
the eigenvalues of $\tilde{K}^{a(b)}_\ell$ are not a discrete value $\pm 1$, but fractional
in general, depending on the values of $g$ and $\lambda_{\ell}$. 
Hence, $\tilde{K}^{a(b)}_{\ell}|\psi_{k}\rangle =I^k_{a(b),\ell}|\psi_{k}\rangle$, where 
$I^{k}_{a(b),\ell}\in \mathbb{R}$ and $|\psi_{k}\rangle$ is $k$-th eigenstate of $H$.
[We use the ascendant order for the eigenstate energies.] 

Here, we should comment on the differences between the form of the LIOMs of 
Eq.(6) and the compact localized state in the Creutz and diamond ladder 
systems \cite{KOI2020,Danieli_1,Roy}. 
The operator of the LIOMs is a multiple spin operator, different from that of 
the compact localized states appearing in the Creutz and diamond ladder systems.
Also, the form of the LIOMs of Eq.(6) in our model is not constrained by the notion of 
the total particle number, the total sum of eigenvalues of the LIOMs is not constrained.
In addition, we remark that we can extend systems to the one with long-range interactions
with modified LIOMs \textit{beyond compact-support},
which we comment on in conclusion. 
In any case, to find a novel Hamiltonian totally given by LIOMs is not an easy task.
A strategy to find a quantum system with an extensive LIOMs is the following;
seek novel LIOMs first and then construct Hamiltonian by using them, which is of physical interest.
[See an example, Ref.~\cite{bilayer}]

In the following sections, we show how the LIOMs label all eigenstates. 
Due to the presence of the extensive number of the LIOMs [${\tilde K}^{a(b)}_{\ell}$], 
the system can exhibit some characteristic dynamics originated from the integrability; some localization phenomenon can occur. 
Furthermore, as characteristic dynamical properties, 
non- or slow-thermalized dynamics emerges \cite{Rigol2007}.

Here, let us remark on significance of the above findings from the view point
of quantum information theory.
As we briefly mentioned in introduction, localization of quantum many-body systems
is currently studied by using knowledge of quantum information theory \cite{Huse2020,Fisher2021,Lavasani2021}.
In quantum error-correction code, notion of stabilizer plays a central role, and compact-support LIOMs are nothing but locally-defined stabilizers. 
Ordinary stabilizers are an element of Pauli group, i.e., a string of Pauli matrices.
The modified LIOMs in Eq.~(\ref{LIOM2}) can be regarded as a new type of stabilizers
as they are linear sums of Pauli strings.
This spatial structure may make codes produced by the modified LIOMs robust against errors.

Before going into detailed analysis and numerical demonstration for the modified LIOMs, 
we would like to discuss a Majorana-fermion representation of the system. 
[Readers who are particularly interested in numerical studies can immediately skip to Sec.~IV and beyond.]

For the non-interacting case ($g=0$), this representation can be obtained rather straightforwardly.
We note that the operators $K^\pm_j$ satisfy the commutation relations similar to 
the hard-core boson creation/annihilation
operators such as, 
\be
(K^+_j)^\dagger = K^-_j, \; (K^\pm_j)^2=0, \; K^+_jK^-_j+K^-_jK^+_j=1,
\label{comKK}
\ee
and also $K^+_j K^-_j={1 \over 2}+{1 \over 2}K_j$ and $[K_j, K^\pm_j]=\pm 2 K^\pm_j$.
From the above properties, we define operators $\chi^\alpha_j \ (\alpha=1,2)$:
\be
\chi^1_j= K^+_j+K^-_j, \;\; \chi^2_j={1 \over i}(K^+_j-K^-_j),
\label{KK}
\ee
which are hard-core bosons satisfying $(\chi^\alpha_j)^2=1$ and $\chi^1_j\chi^2_j+\chi^2_j\chi^1_j=0$.
In order to produce operators from $\{\chi^\alpha_j\}$ that anti-commute 
with each other at different lattice sites, 
we use a Jordan-Wigner transformation such as, 
$K^{\pm}_j \to e^{\pm i \pi \sum_{i<j}{1 \over 2}(K_i+1)} K^\pm_j$, 
and define Majorana fermions, $\{\tilde{\chi}^\alpha_j\}$ as in Eq.~(\ref{KK}).
It is easily verified that $\{\tilde{\chi}^\alpha_j\}$ are Majorana fermions.
In terms of $\{\tilde{\chi}^\alpha_j\}$, $K_j =-i\tilde{\chi}^1_j\tilde{\chi}^2_j$,
and therefore,
\be
H_0=-i \sum_\ell \lambda_\ell \Big[\tilde{\chi}^1_{2\ell}\tilde{\chi}^2_{2\ell}
-\tilde{\chi}^1_{2\ell+1}\tilde{\chi}^2_{2\ell+1}\Big].
\label{H0M}
\ee
Then, the system $H_0$ is expressed in terms of the non-interacting Majorana fermions. 
From this form, it is clear that the system $H_0$ is integrable and 
the Majorana fermions are paired on the site $2\ell$ and $2\ell+1$ 
with an energy given by the disorder $\lambda_\ell$ and do not move.

Let us turn to the interacting case with $g\neq 0$, and focus on terms in $H$ of Eq.~(\ref{Model_mod}) for $j=2\ell$ and $2\ell+1$.
We first define operators 
\be
&&\hat{\alpha}^+_1={2\hat{g} \over [4\hat{g}^2+(2\lambda_\ell - \hat{\epsilon})^2]^{1/2}},\:\:\: 
\hat{\alpha}^+_2={\hat{\epsilon}-2\lambda_\ell \over [4\hat{g}^2+(2\lambda_\ell - \hat{\epsilon})^2]^{1/2}}, \nonumber \\
&&\hat{\alpha}^-_1={2\hat{g} \over [4\hat{g}^2+(2\lambda_\ell + \hat{\epsilon})^2]^{1/2}}, \:\:\:
\hat{\alpha}^-_2=-{\hat{\epsilon}+2\lambda_\ell \over [4\hat{g}^2+(2\lambda_\ell + \hat{\epsilon})^2]^{1/2}},
\label{alphaop}
\ee
where $\hat{g}\equiv gN_{\ell-1}$ and $\hat{\epsilon}\equiv 2[\lambda_\ell^2+\hat{g}^2]^{1/2}$.
We note that $\hat{g}$ and $\hat{\epsilon}$ are operators but they commute with $K^\pm_{2\ell(2\ell+1)}$, and then, 
they can be treated as c-numbers when we study the $j=2\ell/(2\ell+1)$ system. We shall comment on this point later on. 
By using the above-defined operators, we introduce the following operators:
\be
&&\bar{K}^+_{2\ell}=\hat{\alpha}^+_1K^+_{2\ell}+\hat{\alpha}^+_2K^+_{2\ell+1}, \:\:\:
\bar{K}^-_{2\ell}=\hat{\alpha}^+_1K^-_{2\ell}+\hat{\alpha}^+_2K^-_{2\ell+1},  \nonumber \\
&&\bar{K}^+_{2\ell+1}=\hat{\alpha}^-_1K^+_{2\ell}+\hat{\alpha}^-_2K^+_{2\ell+1},  \:\:\:
\bar{K}^-_{2\ell+1}=\hat{\alpha}^-_1K^-_{2\ell}+\hat{\alpha}^-_2K^-_{2\ell+1}.
\label{barK}
\ee
By using Eqs.~(\ref{alphaop}), it is verified that $\{\bar{K}\}$'s in Eq.~(\ref{barK}) satisfy the same
commutation relations with $\{K\}$'s for $j=2\ell/(2\ell+1)$. 
By substituting Eqs.~(\ref{alphaop}) into Eqs.~(\ref{barK}), 
we obtain,
\begin{eqnarray}
&&\bar{K}^+_{2\ell}\bar{K}^-_{2\ell} - \bar{K}^+_{2\ell+1}\bar{K}^-_{2\ell+1}\nonumber\\
&=&{2\lambda_\ell \over \hat{\epsilon}}\Big[K^+_{2\ell}K^-_{2\ell}-K^+_{2\ell+1}K^-_{2\ell+1}\Big] 
+{2\hat{g} \over \hat{\epsilon}}\Big[K^+_{2\ell}K^-_{2\ell+1}+K^-_{2\ell}K^+_{2\ell+1}\Big].
\label{barKKH}
\end{eqnarray}
We also note $N_\ell = 2\bar{K}^+_{2\ell}\bar{K}^-_{2\ell}+2\bar{K}^+_{2\ell+1}\bar{K}^-_{2\ell+1}-2$.
Then, by using Eq.~(\ref{barKKH}), the Hamiltonian $H$ of Eq.~(\ref{Model_mod}) 
is written by
\be
H=\sum_\ell {\hat{\epsilon}_\ell \over 2}
\Big[\bar{K}^+_{2\ell}\bar{K}^-_{2\ell} - \bar{K}^+_{2\ell+1}\bar{K}^-_{2\ell+1}\Big],
\label{HKs}
\ee
where we have returned the suffix $\ell$, $\hat{\epsilon}_\ell \equiv 2[\lambda_\ell^2+(gN_{\ell-1})^2]^{1/2}$.
From the above study of the Hamiltonian $H$, we can introduce quasi-Majorana operators straightforwardly
as in the non-interaction case discussed above.
That is, 
\be
&&\bar{\chi}^1_j= e^{ i \pi \sum_{i<j}{1 \over 2}(\bar{K}_i+1)} \bar{K}^+_j+ 
 e^{-i \pi \sum_{i<j}{1 \over 2}(\bar{K}_i+1)} \bar{K}^-_j, \nonumber  \\
&&\bar{\chi}^2_j={1 \over i}( e^{i \pi \sum_{i<j}{1 \over 2}(\bar{K}_i+1)} \bar{K}^+_j
- e^{- i \pi \sum_{i<j}{1 \over 2}(\bar{K}_i+1)} \bar{K}^-_j), \nonumber
\ee
and
\be
H=-i \sum_\ell {\hat{\epsilon}_\ell \over 2}  \Big[\bar{\chi}^1_{2\ell}\bar{\chi}^2_{2\ell}
-\bar{\chi}^1_{2\ell+1}\bar{\chi}^2_{2\ell+1}\Big].
\label{HM}
\ee
Contrary to the system described with $H_0$,
the system $H$ in Eq.~(\ref{HKs}) contains interactions between $\{\bar{\chi}\}$'s located on $(2\ell-2, 2\ell-1)$ sites
and those on $(2\ell,2\ell+1)$ sites.
However terms on the right-hand side of Eq.~(\ref{HKs}) commute with each other, and physical Hilbert space
is divided into subsectors with definite values of $\{N_\ell\}$.
Energy eigenvectors and eigenvalues are obtained in each subsector rather straightforwardly. 
As a result, from the form of Eq.~(\ref{HM}), 
even in the interacting case ($g\neq 0$), the system can be written by a decoupled form of quasi-Majorana pairs without any ‘quasi-Majorana hopping’. This representation gives an insight that the system is integrable and exhibits some specific localized phenomena, which we study below. Furthermore, the above hard-core boson as well as quasi-Majorana fermion representation
play an important role when we later study effects of the Ising-type interactions,
which are nothing but hopping of these particles. 
It should be noted that from the form of Eq.~(\ref{HM}), the Anderson localization does not appear strictly due to the absence of some exponentially decay Anderson orbital. 
But, localization-like phenomenon occurs from the integrability and the locality of ${\bar K}^{+}_j{\bar K}^{-}_j$, which is compact-support. 
If one adds some weak interactions breaking the integrability, slow-thermalization dynamics can occur, analogous to that in the ordinary MBL (We numerically investigate it later). Even though the derivation of the model of Eq.~(\ref{HM}) is somewhat complex, the integrability and locality can give some interesting insights in future research on many-body quantum systems and quantum information theory.

Here, we further comment on the use of the nomenclature `quasi-Majorana'.
As explained in the above, $H$ in Eq.~(\ref{HKs}) is expressed in terms of
`quasi-Majorana fermions' $\{\bar{\chi}\}$'s.
However, $\{\bar{\chi}\}$'s are {\it not} genuine fermions.
As the coefficient $\{\alpha\}$'s in Eq.~(\ref{alphaop}) are operators, and as a result, e.g., $\bar{K}^\pm_{2\ell}$
and $\bar{K}^\pm_{2\ell-1}$ do not commute with each other, and therefore the obtained 
$\{\bar{\chi}\}$'s for $j=2\ell$ and $j=2\ell-1$
do not anti-commute with each other.
This is not remedied by a simple Jordan-Wigner-type transformation although this flaw 
does not matter
unless hopping terms between $2\ell$ and $2\ell-1$ sites, such as
$\bar{K}^{+}_{2\ell-1}\bar{K}^{-}_{2\ell}$, are included in the system Hamiltonian. Because of its complicated commutation relations, quasi-Majorana
representation cannot be used for large system-size calculations such as
quench dynamics, unfortunately.

%%%%%%%%%%%%%%%%%%%%%%%%%%%%%%%%%%%%

\section{Study of small system}\label{SecIV}
We study a small system with $L=6$ to see how the system is affected by the interactions,
and how the eigenstates are characterized by the LIOMs. It is important to
confirm the presence of the LIOMs numerically due to the complexity of the model, 
even though the explicit form of the LIOMs in Eq.(6) is analytically given. 

In this small system, we first focus on the $\ell$-th unit-cell in the stabilizer lattice in Fig.~\ref{Fig0} 
and investigate the eigenvalues and eigenstates of the non-interacting Hamiltonian $H_0$. 
For any $\lambda_\ell$, there are three energy levels, 
$\epsilon=-2\lambda_\ell$, $0$, $2\lambda_\ell$, 
where their eigenstates are given by $|\bar{1}1\rangle_{\ell}$ for $\epsilon=-2\lambda_j$, two orthogonal 
linear-superposed states denoted by 
$\alpha|11\rangle_{\ell}+\beta |\bar{1}\bar{1}\rangle_{\ell}$ for 
$\epsilon=0$, and $|1\bar{1}\rangle$ for $\epsilon=2\lambda_j$, where we have introduced 
a notation of cluster-based state such as 
$|1\bar{1}\rangle_{\ell}=K^{a+}_{\ell}K^{b-}_{\ell}|\Uparrow\rangle_{\ell}$ ($|\Uparrow\rangle_{\ell}$ is all up states around 
$\ell$ unit-cells in the original spin lattice). 
Therefore, $K^a_\ell |1{\bar 1}\rangle_\ell=|1\bar{1}\rangle_\ell$, $K^b_\ell|1{\bar 1}\rangle_\ell = - |1{\bar 1}\rangle_\ell$. 
The presence of an arbitrary linear-superposed state at zero energy of $H_0$ prevents
unique-labeling of eigenstates by the LIOMs. 
A pair of the doubly-degenerate states can be chosen arbitrarily 
as long as they are orthogonal to each other.
Thus, we take $|\bar{1}\bar{1}\rangle_{\ell}$ and $|11\rangle_{\ell}$ as a pair of orthogonal eigenstates 
with zero energy of $H_{0}$. 
To impose this choice in practical calculations, 
we introduce very small random potential, $\sum_{\ell}\delta h_{\ell}[K^{a}_{\ell}+K^{b}_{\ell}]$ with 
$\delta h_{\ell}\in [-\delta h,\delta h]$, $\delta h=0.5 \times 10^{-5}W$. 
This `fictitious' disorder gives little effect to the entire physics, especially, to the dynamical behavior of the system.  
Under this manipulation, ``cluster-basis'' eigenstates of $H_0$ are described as 
$|\psi(\{a_{\ell}\},\{b_{\ell}\})\rangle = \prod^{L/2}_{\ell=0}K^{a\:a_{\ell}} _{\ell}K^{b \:b_{\ell}}_{\ell}|\Uparrow\rangle$
\cite{Jeyaretnam}, where $L$ is an even number, the sets of $\{a_{\ell}\}, \{b_{\ell}\}$ are a sequence of $+$ 
and $-$ labels and $|\Uparrow\rangle$ is a $L$-site ferromagnetic state with all spins up in the original spin basis.

Removing the degeneracy of the zero-energy state of $H_0$ in a single unit-cell as 
explained in the above, we study
the effects of the interaction $H_{\rm int}$ on the cluster-basis eigenstates of $H_{0}$. 
Here, the interaction term $V_{\ell}$ acts over two unit-cells as shown in the lower 
lattice in Fig.~\ref{Fig0}. 
Only four cluster-basis states on the $\ell -1$ and $\ell$ unit-cells are changed: 
(i) $| 11 \rangle_{\ell -1}| 1\bar{1} \rangle_{\ell}$, 
(ii) $| 11 \rangle_{ \ell -1}|\bar{1}1\rangle_{\ell}$,
(iii) $|\bar{1}\bar{1}\rangle_{\ell -1}|1\bar{1}\rangle_{\ell}$, 
(iv) $|\bar{1}\bar{1}\rangle_{\ell -1}|\bar{1}1\rangle_{\ell}$. 
The interaction $V_{\ell}$ mixes (i) and (ii) ((iii) and (iv)), and creates a superposed state of
$|1\bar{1}\rangle_{\ell}$ and $|\bar{1}1\rangle_{\ell}$. 
For the other cluster-basis states
on the two unit-cells are a null state of $V_{\ell}$. 

From the above observation about the action of $V_{\ell}$, 
characteristic eigenstates for the $L=6$ interacting system of $H$ are obtained straightforwardly. 
As an examples,  
$$
|\psi^{L=6}_{1,\pm}\rangle=|11\rangle_{0}|s^{\pm}\rangle_{1}|11\rangle_{2}
$$ 
where
$$
|s^{\pm}\rangle_{1}=\alpha^{\pm}_{1}|1\bar{1}\rangle_{1}+\alpha^{\pm}_{2}|\bar{1}1\rangle_{1}
$$ 
and 
$
(\alpha^{\pm}_{1},\alpha^{\pm}_{2})=[4g^2+(2\lambda_{1}-\epsilon^{\pm}_{1})^2]^{-1/2}(2g, \epsilon^{\pm}_{1}-2\lambda_{1})
$
with $\epsilon^{\pm}_{1}=\pm 2[\lambda^2_{1}+g^2]^{1/2}$.
The state, $|\psi^{L=6}_{1,\pm}\rangle$, is an eigenstate for all LIOMs, 
$\tilde{K}^{a(b)}_{\ell}$, with integer or fractional eigenvalues, e.g., 
$$
\tilde{K}^{a}_{1}|\psi^{L=6}_{1,\pm}\rangle=\frac{\epsilon^{\pm}_{1}}{2\lambda_{1}}|\psi^{L=6}_{1,\pm}\rangle,  \;\;
\tilde{K}^{b}_{1}|\psi^{L=6}_{1,\pm}\rangle=-\frac{\epsilon^{\pm}_{1}}{2\lambda_{1}}|\psi^{L=6}_{1,\pm}\rangle.
$$
From this observation of $|\psi^{L=6}_{1,\pm}\rangle$, 
certain cluster-basis eigenstates of $H_0$, $|\psi(\{a_{\ell}\},\{b_{\ell}\})\rangle$ are mixed by the interactions 
$H_{\rm int}$, 
however, the mixing is local and only small numbers of cluster-basis eigenstates are affected. 
These observations indicate that global hybridization does not occur by the interactions $H_{\rm int}$ 
due to the presence of the extensive number of the LIOMs, $\tilde{K}^{a(b)}_{\ell}$.
[We give observation of the general structure of the Hilbert space for the $L$-site 
system in Appendix A.]
Following the above analytical observation, we numerically verify the presence of the LIOMs in the $L=6$ system. 

In all numerical calculations in this work, we employ the Quspin solver \cite{Quspin},
where the spin Hamiltonian matrix including any multiple-body spin interactions can be
constructed and numerical exact diagonalization is carried out with some efficient 
Python packages. In this work, we employ periodic boundary condition.

We observe whether the operators $\{\tilde{K}^{a}_{j}\}$ operate as the LIOMs for all
eigenstates, i.e., 
all energy eigenstates are the eigenstates of the LIOMs.
To this end, we first examine whether the resulting states obtained by acting $\tilde{K}^{a}_{\ell_0}$ 
on energy eigenstates are proportional to the original ones. 
Numerically, for $k$-th eigenstate $|\psi_{k}\rangle$, we define 
$|\tilde{\psi}_{k}\rangle\equiv \tilde{K}^{a}_{\ell_0}|\psi_{k}\rangle/|\langle \psi_{k}|(\tilde{K}^{a}_{\ell_0})^{\dagger}\tilde{K}^{a}_{\ell_0}|\psi_{k}\rangle|$ 
and calculate $|\Delta \psi_{k}|^2=||\psi_{k}\rangle-|\tilde{\psi}_{k}\rangle|^2$. 
Then, if $|\Delta \psi_{k}|^2=0$ or $4$, 
$|\psi_{k}\rangle$ is also an eigenstate of $\tilde{K}^{a}_{\ell_0}$ with the eigenvalue, 
$I^k_{a,\ell_0}=\langle \psi_{k}|\tilde{K}^{a}_{\ell_0}|\psi_{k}\rangle$. 
Here, we set $\ell_0=1$ for the practical calculation.

\begin{figure*}[t]
\begin{center} 
\includegraphics[width=15cm]{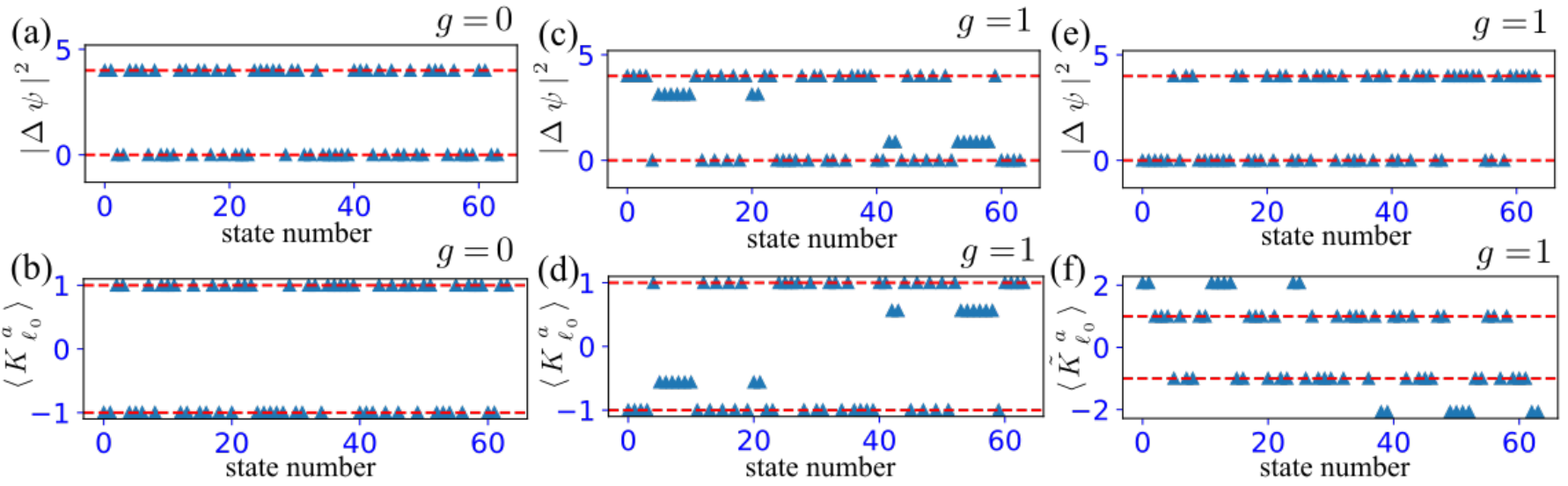}  
\end{center} 
\caption{Identification of eigenstates and their LIOM-eigenvalues for three unit-cell system ($L=6$). 
(a) $|\Delta \psi_k|^2$ for each eigenstate for $g=0$ case. 
(b) Each eigenvalue of $K^{a}_{\ell_0}$ for $g=0$ case. 
(c) $|\Delta \psi_k|^2$ for each eigenstate for finite $g$ in terms of $K^{a}_{\ell_0}$. 
(d) Each eigenvalue $\langle \psi_{k}|K^{a}_{\ell_0}|\psi_{k}\rangle$ for finite $g$ case. 
(e) $|\Delta \psi_k|^2$ for each eigenstate for finite $g$ in terms of $\tilde{K}^{a}_{\ell_0}$. 
(f) Each eigenvalue of $\tilde{K}^{a}_{\ell_0}$ for finite $g$ case. 
All results are obtained for a single-shot disorder.
Energy eigenstates are numbered in the ascendant order.}
\label{Fig1}
\end{figure*}

We show numerical results where we set $W=2$. 
For the $g=0$ case, all eigenstates for single-shot disorder realization are labeled by 
the original stabilizers, $\{K^{a(b)}_{\ell}\}$.  
Each $|\Delta \psi_{k}|^2$ takes $0$ or $4$ and the LIOMs' eigenvalues are $I^k_{a,1}=\pm 1$ 
as shown in Figs.~\ref{Fig1} (a) and (b). 
For the $g=1$ case, we first operate the original $\{K^{a(b)}_\ell\}$ to energy eigenstates for $g=1$
to see that $\{K^{a(b)}_\ell\}$ are \textit{not} genuine LIOMs.
As shown in Figs.~\ref{Fig1} (c) and (d),
some of $|\Delta \psi_{k}|^2$ deviate from $0/4$ and $\langle \psi_{k}|\tilde{K}^{a}_{\ell_0}|\psi_{k}\rangle$ 
deviate from $\pm 1$. 
On the other hand for the operation of $\tilde{K}^{a(b)}_{\ell_0}$, as shown in Fig.~\ref{Fig1} (e) 
$|\Delta \psi_{k}|^2$ takes $0$ or $4$, that is, $\tilde{K}^{a}_{\ell_0}$ is a genuine LIOM with eigenvalues 
$\{I^{k}_{a,1}\}$, where some of $I^{k}_{a,1}$ takes a disorder-dependent fractional value (See Fig.~\ref{Fig1} (f)). 
Also, note that the interactions $H_{\rm int}$ have some 
large kernel space, $H_{\rm int}|\psi_{k}\rangle=0$, 
which means that a substantial number of eigenstates still have $\langle\tilde{K}^{a}_{\ell}\rangle =\langle K^{a}_{\ell}\rangle$ 
with eigenvalues $\pm 1$ [Fig.~\ref{Fig1} (f)]. 

From the above study on the small system, we found that
the modified LIOMs, $\{{\tilde K}^{a(b)}_{\ell}\}$, indeed label all energy eigenstates.
We expect that this holds for the system with larger sizes, see Appendix B, where we show numerical verification
of this expectation for a larger system. 
Also, the interaction $H_{\rm int}$ mixes only a small number of local cluster-basis eigenstates of $H_0$. 
This fact implies that the interacting model in the present work exhibits some characteristic dynamics originated from the integrability. 
In what follows, to elucidate it we numerically investigate the dynamics of the system.

%%%%%%%%%%%%%%%%%%%%%%%%%%%%%%%%%%%% 
%%%%%%%%%%%%%%%%%%%%%%%%%%%%%%%%%%%%
%%%%%%%%%%%%%%%%%%%%%%%%%%%%%%%%%%%%%%%%%%%%%%%%%%
\section{Numerical demonstration of characteristic dynamics}\label{SecV}
In this section, we numerically investigate quench dynamics for the system $H_{0}+H_{int}$ where we set $W=2$. 
In the previous section, we verified that the system is indeed integrable even in the presence of the interaction $H_{int}$, 
where the modified LIOMs label all eigenstates, i.e., act as local conserved quantities. 

In the following numerical calculations of time evolution of the system, 
we employ exact diagonalization \cite{Quspin}, where the accessible system size is up 
to $L=16$.  
In what follows, we remove the `fictitious' disorder, $\delta h_{\ell}=0$. 
The interaction $H_{\rm int}$ 
affects to the system locally, mixing a few cluster-basis eigenstates, and has large kernel space, 
$H_{\rm int}|\psi_{k}\rangle=0$. 
From these facts and the existence of the extensive number of the LIOMs, $\tilde{K}^{a(b)}_\ell$, 
we expect certain characteristic dynamical phenomena in the interacting system [$H$ in Eq.~(\ref{Model_mod})], especially, 
the quench dynamics that exhibits non- or slow-thermalization \cite{Rigol2007}. 
To observe this expectation, we employ the return probability, given by 
$$
\langle{RP}\rangle= |\langle \psi(t)|\psi(0)\rangle|^2,
$$ 
where $|\psi(t)\rangle$ is the many-body wave function at time $t$, and the EE, defined as
$$
S=-\mbox{Tr} [\rho_A \ln (\rho_A)],
$$ 
with $A$-subsystem reduced density matrix 
$\rho_A=\mbox{Tr}_B[\rho]$, where $\rho$ is a density matrix of the entire system and the subsystem is set to $L/2(L/2+1)$-site system for even (odd) $L$. 
For the practical calculation, $A$ and $B$ subsystems are set as shown in 
Fig.~\ref{Fig0}. 
In what follows, time is measured in units $[g/\hbar]$ and 
we set a random cluster-basis state (e.g.,$|1\bar{1}1\bar{1}\bar{1}\cdots \rangle$) as an initial state which was introduced in Sec.4. 
The random cluster-basis state is short-range entangled, and the value of the EE is obtained by cutting two cluster states. This value is an initial value of the EE in the quench dynamics. 
In the quench dynamics, we average over 60 samples for the initial state and disorder realizations, 
$\{\lambda_{\ell}\}$.

\begin{figure}[t]
\begin{center} 
\includegraphics[width=10cm]{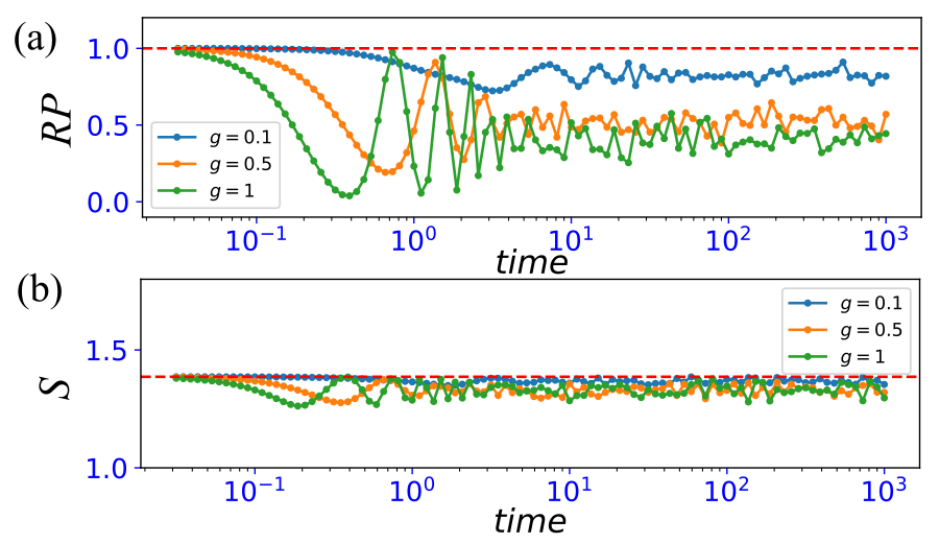}  
\end{center} 
\caption{Dynamics of the return probability [(a)] and entanglement entropy [(b)]. 
We set $L=12$ and averaged over $60$ disorder realization for $W=2$ and initial random cluster-basis state.
The entanglement entropy is quite stable, and its value is close to $2\ln 2$ 
(corresponding to the value obtained by cutting two clusters).
We set $\delta h=0$.}
\label{Fig2}
\end{figure}

Numerical results are displayed in Fig.~\ref{Fig2}. 
The return probability remains large finite values for a long period and the EE remains low values around 
$2\ln 2$ \cite{clusterEE} as shown in Figs.~\ref{Fig2} (a) and ~\ref{Fig2} (b), 
where we set $\delta h=0$. 
These behaviors retain for large $g$'s. 
The results of the return probability in Fig.~\ref{Fig2} (a) indicate that initial-state
information is preserved, which means ergodicity breaking 
and also the behavior of the EE implies all eigenstates of $H$ remain to 
be low-entangled. 
These numerical results show that even for the cluster model with the interactions 
$H_{\rm int}$, ergodicity breaking dynamics with short-range entanglement is retained. 
This may seem a little bit odd as the Hamiltonian in Eq.~(\ref{HKs}) 
has a standard form of LIOMs systems [observed by expanding $\hat{\epsilon}_\ell$
in powers of $g$], and then log-like evolution of the EE may be expected.
The origin of this unconventional behavior of the EE comes from the existence of 
the extensive number of commuting operators $\{\tilde{K}^{a(b)}_\ell\}$, and
the Hamiltonian is given by a linear combination of them.
This is very specific nature of the present system.

%%%%%%%%%%%%%%%%%%%%%%%%%%%%%%%%%%%% 
%%%%%%%%%%%%%%%%%%%%%%%%%%%%%%%%%%%%
In addition, we show a numerical estimation of the effect of the fictitious
disorder in Appendix C 
and also the system-size dependence of the return probability in Appendix D. 
These results indicate that small fictitious disorder $\delta h_{\ell}$ does not have 
a significant effect on dynamics 
and initial-state information is preserved even for large system sizes.

%%%%%%%%%%%%%%%%%%%%%%%%%%%%%%%%%%%%%%%%%%%%%%
\section{Effects of additional interactions: Many-body localization dynamics}\label{SecVI}

In this section, we shall examine the stability and robustness 
of the characteristic ergodicity breaking dynamics originated from the integrability 
studied in the previous sections.
To this end, we investigate the effects of the Ising-type interactions
as a typical perturbation respecting the symmetries of the system.
The motivation of this study comes from the seminal research concerning the stability
of Anderson localization against interactions \cite{Basko2006}. 
The Ising interaction is given by 
$V_{zz}=v_{zz}\sum^{L-1}_{j=0}\sigma^{z}_{j}\sigma^{z}_{j+1}$, where $v_{zz}$ is a
controllable parameter. 
We numerically observe how the ergodicity breaking properties of the model [Eq.~(\ref{Model_mod})] change.
Obviously for finite $v_{zz}$, the operators $\{\tilde{K}^{a(b)}_{\ell}\}$ are no longer exact LIOMs. 

To obtain physical insight of the system $H+V_{zz}$,
it is quite helpful to use the quasi-Majorana (hard-core-boson) representation 
introduced in Sec.~III.
As $\sigma^z_j=K^+_j+K^-_j=\chi^1_j$,  we have
$V_{zz}=v_{zz}\sum^{L-1}_{j=0}\chi^1_j\chi^1_{j+1}$.
Therefore, $H_0$ in Eq.~(\ref{H0M}) plays a role of random potentials, whereas $V_{zz}$
generates hopping amplitudes of the Majorana fermions.
We expect that the system $H_0+V_{zz}$ exhibits typical phenomena of Anderson localization for finite values of $v_{zz}$,
and adding the interactions $H_{\rm int}$ to $H_0+V_{zz}$ induces ordinary MBL in $H+V_{zz}$.
Therefore, we expect that $V_{zz}$ play a role of the leading term in $H+V_{zz}$, and  
it induces somewhat different localization properties from the system with $V_{zz}=0$,
in particular the EE. 
This is a very interesting property of the present system
exhibiting crossover from the compact localization-like phenomenon to Anderson-based MBL.
Studying this crossover is an interesting future problem. 
In addition, the interaction, $V_{zz}$, preserves the 
$\mathbb{Z}_{2}\times \mathbb{Z}^{T}_{2}$ symmetry,
and therefore, we can also expect that the SPT order persists even in the presence of
$V_{zz}$. 

To verify the above expectations, we first carried out the level-spacing analysis
for the system with finite $v_{zz}$ as shown in Appendix E, and we found the results 
indicating the localization tendency of the system with $V_{zz}$.
Then, we numerically investigate the quench dynamics of the model with finite $g$ and $v_{zz}$, 
where we set the cluster-basis Neel state $|1\bar{1}1\bar{1}1\cdots\rangle$ as an initial state and observe the return probability and EE. 
The cluster-basis state is also short-range entangled as in the random cluster-basis state employed in the calculation in Fig.~\ref{Fig2}. 
The initial value of the EE is finite in the quench dynamics. 
In the conventional cluster spin model, to characterize the bulk SPT order, a string order parameter is employed \cite{Son,Smacchia}. 
We also use a similar quantity, transformed into a periodic form. 
It is a loop order defined by 
$$
\langle {LO}\rangle=\langle \Psi(t)|{\hat L}|\Psi(t)\rangle,
$$ 
with ${\hat L}=\prod^{L/2-1}_{\ell=0}K^{a}_{\ell}$. 
We expect that the loop order diagnoses the presence of the SPT order. 
Note that here the initial state is different from that in the previous calculation in Fig.~\ref{Fig2} (a) and ~\ref{Fig2} (b). 
This initial state is similar to a typical initial state that is composed of Pauli-spin eigenstates and is used in the study of quench dynamics of the conventional MBL systems~\cite{Abanin,Bardarson} and can be easily prepared experimentally \cite{Abanin}.

%%%%%%%%%%%%%%%%%%%%%%%%%%%%%%%%%%%% 
\begin{figure}[t]
\begin{center} 
\includegraphics[width=10cm]{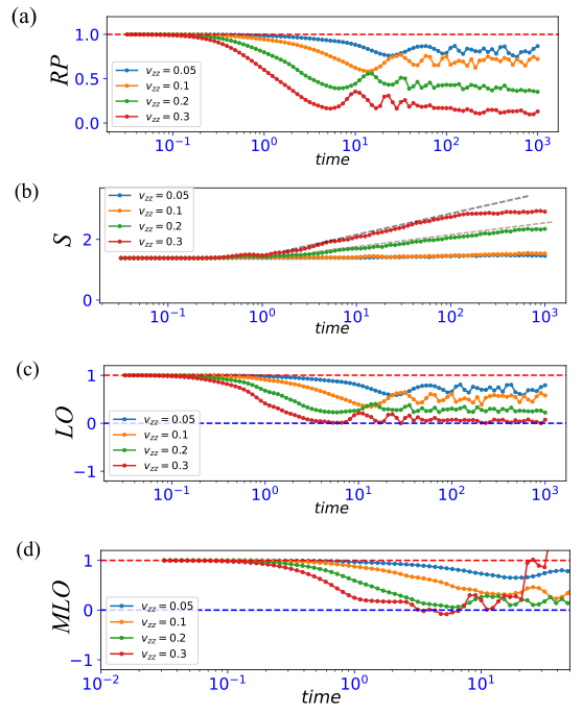}  
\end{center} 
\caption{
Dynamics of the return probability [(a)], entanglement entropy [(b)], loop order [(c)] and modified loop order [(d)]. 
We set $L=12$, $W=2$, $g=1$ and $\delta h=0$ and averaged over $40$ disorder realizations.
Note that the behavior of the return probability for $v_{xx}=0.05$ is fairly different 
from that for $g=1$ in Fig.~\ref{Fig2} (a). The reason comes from the different 
choice of initial state.}
\label{Fig3}
\end{figure}
%%%%%%%%%%%%%%%%%%%%%%%%%%%%%%%%%%%%
Numerical results are shown in Fig.~\ref{Fig3}, where we set $g=1$. For small $v_{zz}$,
the value of the return probability remains finite [Fig.~\ref{Fig3} (a)], the increase of 
the EE is much
suppressed [Fig.~\ref{Fig3} (b)] and also the value of the loop order remains finite
[Fig.~\ref{Fig3} (c)] for a long period. 
Obviously, these are \textit{conventional MBL behavior with the SPT order}.
For larger $v_{zz}=0.3$, localization tendency is weakened, i.e.,
the increase of the EE is enhanced with logarithmic growth and saturates with larger 
values, approaching the Page value \cite{Page}, $(L\log 2-1)/2$. 
Correspondingly, the values of the return probability and loop order also are decreasing, i.e.,
thermalization tendency is enhanced and the SPT order is fading away. 
In addition, we investigated the behavior of a modified loop order (MLO) described by $\tilde{K}^{a}_{\ell}$, 
defined by 
$$
\langle{MLO}\rangle = \langle \Psi(t)|\biggl[\prod^{L/2-1}_{\ell=0}\tilde{K}^{a}_{\ell}\biggr]|\Psi(t)\rangle. 
$$
Note that the operator depends on the set of disorder $\{\lambda_{\ell}\}$. 
We set the cluster-basis Neel state $|1\bar{1}1\bar{1}1\cdots\rangle$ 
as an initial state, 
which has $\langle{MLO}\rangle =1 (-1)$ for an even (odd) $L/2$. 
The dynamics of the MLO for various $v_{zz}$ is shown in Fig.~\ref{Fig3} (d). 
For early times, the MLO sustains the initial values. 
The MLO starts to decay after a long period in the time evolution,
with large oscillations since state mixing with large LIOM eigenvalues occurs 
in the process of long-time evolution.  
More precisely, this large oscillation comes from the instability of the cluster-basis Neel state (the initial state) in the dynamics. 
The initial cluster-basis Neel state has the eigenvalue $+1$ for all $\{\tilde{K}^{a}_{\ell}\}$'s in the MLO operator, 
and then it has $\langle MLO\rangle=1$. 
Under the time evolution, the initial state is deformed by $H_{\rm int}$ as well as the Ising interaction, $V_{zz}$, and the latter effect is larger for a larger value of $v_{zz}$. As the result, the $g$-dependent terms in $\{\tilde{K}^{a}_{\ell}\}$ start to operate to generate nontrivial behavior of the MLO [See the $v_{zz}=0.3$ case in Fig.~\ref{Fig3} (d)]. 

%%%%%%%%%%%%%%%%%%
%%%%%%%%%%%%%%%%%%%%%%%%%%%%%%%%%%%% 
\begin{figure}[t]
\begin{center} 
\includegraphics[width=12cm]{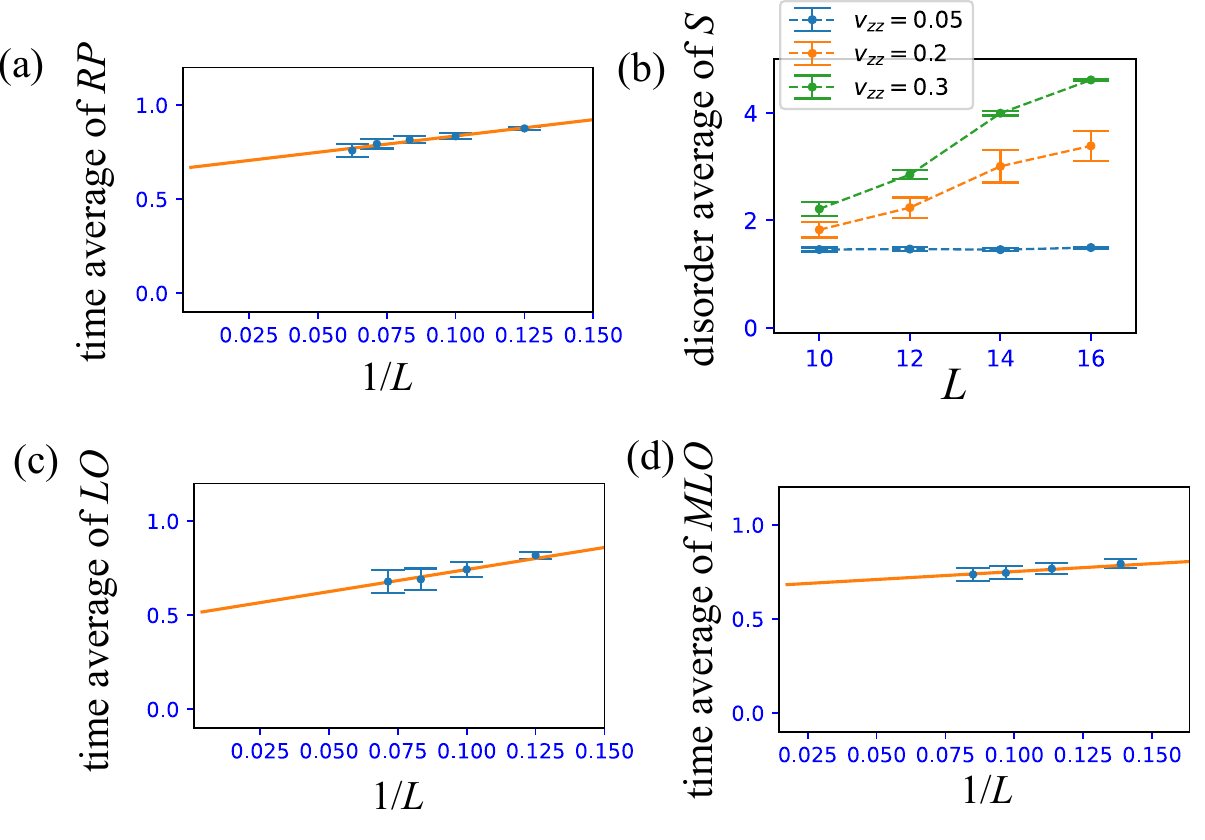}  
\end{center} 
\caption{System size dependence of the time average of the return probability [(a)], loop order [(c)] and modified loop order [(d)] for $v_{zz}=0.1$. We employed the time intervals, $t\in [0.03,10^{3}]$ for the calculation (a), (c) and (d). 
For the data (c) and (d), we take the absolute value of the loop order and MLO for an odd $L/2$.
Panel (b) is the system size dependence of the disorder-averaged saturation value of the (half-chain) entanglement entropy at a time point $t=10^3$. 
For all data, we set $W=2$, $g=1$ and $\delta h=0$.
}
\label{Fig3_2}
\end{figure}
%%%%%%%%%%%%%%%%%%%%%%%%%%%%%%%%%%%%

Finally, we would like to comment on the system-size dependence and thermodynamic limit 
($L \to \infty$) of the return probability, EE, loop order, and modified loop order for the system with $V_{zz}$. In Fig.~\ref{Fig3_2} (a), we display the numerical results of the system-size dependence of the return probability,
and find that the time-averaged return probability only slightly decrease as $L$ increases for the case of $v_{zz} = 0.1$. 
The details of the system-size dependence in the dynamics
are shown in Appendix D.

In Fig.~\ref{Fig3_2} (b), we show a system-size dependence of the disorder averaged saturation value of the EE at time $t=10^{3}$, 
when the EE of each data in Fig.~\ref{Fig3} (b) almost saturates 
[the number of the disorder realization is 100, 80, 60, 40 and 24 for $L=8,10,12,14$ and $16$, respectively]. 
The result indicates that for small $v_{zz}$ the saturation value of the EE seems to exhibit area-law, $S\sim {\rm const.}$ and as increasing $v_{zz}$ the saturation value of the EE increase as $L$ increases, which is a signal of the sub-volume law of the EE (the behavior deviates from the area law, at least). 
This behavior is fairly close to that in the conventional MBL \cite{Nandkishore2015,Bardarson}.

Also, the system-size dependence of the time-averaged loop order and MLO in Fig.~\ref{Fig3_2} (c) and Fig.~\ref{Fig3_2} (d) for $v_{zz} = 0.1$ indicates
that the finite values of them remain even for larger systems. Then the existence of SPT order in the MBL state is indicated in the system with weak $V_{zz}$. 
The detailed dynamics of the loop order and MLO is shown in Appendix D. 
These results support the existence of SPT order in the system $H + V_{zz}$ even for large system sizes. 

However, summarizing the results of the system-size dependence, we expect that the system can get eventually thermalized in the final stage. 
In this sense, the model exhibits the presence of a “slow-thermalization”.

%%%%%%%%%%%%%%%%%%%%%%%%%%%%%%%%%%%
\section{Discussion and conclusion}\label{SecVII}
We showed that a specific type of interacting cluster model with disorder
possesses modified LIOMs. 
Modified LIOMs, which are a counterpart of the stabilizer operators in the original 
cluster spin model, 
were found for arbitrary strength of interactions and disorder. 
We showed that the LIOMs label all energy eigenstate by analytical and  numerical
methods.
The locality of the compact-support LIOMs and the extensive number of them imply some characteristic dynamical phenomena.
We numerically demonstrated that the quench dynamics of the system shows slow-thermalized dynamics 
(ergodicity breaking dynamics), and also the SPT order of the original cluster spin model somewhat preserves by observing the string order.
Furthermore, we considered the effects of the Ising interactions, where the modified 
LIOMs are no longer exact stabilizers.
In the Majorana-fermion picture, the Ising interactions are nothing but its hopping. 
Then, we expect that the system exhibits genuine MBL. 
We numerically verified that non-thermalized dynamics is stable against weak Ising interactions. 
The numerical result is a signature of the presence of the MBL. 
The ergodicity breaking dynamics characterized by the LIOMs [$\tilde{K}^{a(b)}_{\ell}$] survives 
in the presence of the Ising interactions, $V_{zz}$.

Finally, we comment that another type of the LIOMs can be defined, which have not compact but a long-tail support. 
If in the model of Eq.~(\ref{Model_mod}), we change $H_{\rm int}$ to 
$H'_{\rm int}=\sum_{r,\ell}V^{r}_{\ell}$, where 
$V^{r}_{\ell}\equiv g e^{-|r|}N_{\ell-r-1}(K^{a+}_{\ell}K^{b-}_{\ell}+K^{a-}_{\ell}K^{b+}_{\ell})$, 
another type of LIOMs can be constructed such as 
$L^{a}_{\ell}\equiv K^{a}_{\ell}+\frac{1}{2\lambda_{\ell}}\sum_{r}V^{r}_{\ell}$ 
and $L^{b}_{\ell}\equiv K^{b}_{\ell}-\frac{1}{2\lambda_{\ell}}\sum_{r}V^{r}_{\ell}$.
These LIOMs are not compact but have a long-tail support.
The simplest case with only $r=0$ and $1$ terms in $H'_{\rm int}$ can be quickly investigated
numerically in the same way as the numerical calculation in Figs.~\ref{Fig1} (e) and (f). 
The result is shown in Appendix G. 
Surely, we confirmed that in the simplest case the LIOMs also characterize 
all eigenstate in the system.
Hence, this system may induce MBL phenomena, which poses a future work.

%%%%%%%%%%%%%%%%%%%%%%%%%%%%%%%%%%%%%%%%%%%%

%%%%%%%%%%%%%%%%%%%%%%%%%%%%%%%%%%%%%%%%%%%%%%%%%%%

\section*{Acknowledgments}
The work is supported by JSPS
KAKEN-HI Grant Number JP21K13849 (Y.K.). T.O. has been supported by the Program for Developing 
and Supporting the Next-Generation of Innovative Researchers at Hiroshima University.

%%%%%%%%%%%%%%%%%%%%%%%%%%%%%%%%%%%%%%%%%%%%%%%%%%%%%%%%
%\appendix
%\renewcommand{\thefigure}{\Alph{section}.\arabic{figure}}
%\setcounter{figure}{0}
%\renewcommand{\theequation}{A.\arabic{equation}}

\section*{Appendix A. Structure of Hilbert space under `fictitious' disorder}
For $L$-site system, where the total dimension of Hilbert space is given by $N_D=2^{L}$, 
all eigenstates $|\psi_{k}\rangle$ for $H$ of Eq.~(\ref{Model_mod}) with $g\neq 0$ under weak `fictitious' disorder are 
classified into two classes since the interaction acts to only the following four states in two unit-cells: 
$|11\rangle_{\ell-1}|1\bar{1}\rangle_{\ell}$, 
$|11\rangle_{\ell-1}|\bar{1}1\rangle_{\ell}$,
$|\bar{1}\bar{1}\rangle_{\ell-1}|1\bar{1}\rangle_{\ell}$,
and 
$|\bar{1}\bar{1}\rangle_{\ell-1}|\bar{1}1\rangle_{\ell}$.

The first class is composed of eigenstates satisfying 
$H_{\rm int}|\psi_{k}\rangle=g\sum_{\ell}V_{\ell}|\psi_{k}\rangle = 0$, that is, the eigenstate $|\psi_{k}\rangle$
 is a null state for $H_{\rm int}$. 
On the other hand, the second class is composed of the ones satisfying 
$H_{\rm int}|\psi_{k}\rangle= v_{k}|\psi_{k}\rangle$, where $v_{k}$ is finite real value. 
The above classification can be understood by observing how the interaction acts on the eigenstates of $H_{0}$.

In $N_D$ eigenstates of $H$, $2^{L/2+1}$ eigenstates are totally unaffected by the
interaction, $H_{\rm int}$.  
These eigenstates consist of the following two categories: (I) in the eigenstate, 
the state of each unit-cell is given by
 $|11\rangle_{\ell}$ or $|\bar{1}\bar{1}\rangle_{\ell}$ (total $2^{L/2}$ eigenstates). 
(II) in the eigenstate, the state of each unit-cell is given by $|1\bar{1}\rangle_{\ell}$ or $|\bar{1}1\rangle_{\ell}$ (total $2^{L/2}$ eigenstates). Needless to say, these total $2^{L/2+1}\equiv N^{1}_{D}$ eigenstates are eigenstates of $H_{0}$. 

On the other hand, the number of eigenstates of $H$ with finite interaction energies is obtained by 
counting the number of eigenstates of $H_0$ affected by the interaction $H_{\rm int}$. 
This number can be counted as follows: 
(a) We consider the eigenstate in which the state of each unit-cell is given by
$|11\rangle_{\ell}$ or $|\bar{1}\bar{1}\rangle_{\ell}$ (total $2^{L/2}$ eigenstates). 
(b) For each eigenstate, select $k$ unit-cells and change their states of the unit-cells to 
$|1\bar{1}\rangle_{\ell}$ or $|\bar{1}1\rangle_{\ell}$, where $k$ takes from $1$ to $L/2-1$. 
The total number of eigenstates obtained in this way is 
$\sum^{L/2-1}_{k=1}2^{L/2-k} \ _{L/2}C_{k} \ 2^{k} =4^{L/2}-2^{L/2+1}\equiv {N^2}_{D}$, which are affected by 
the interaction $H_{\rm int}$. 
These $N^2_{D}$ eigenstates are the second ones. 
In addition, the total sum of the eigenstates of 
the first and second class is $N^{1}_{D}+N^{2}_{D}=N_D$.

Furthermore, the interaction $H_{\rm int}$ mixes the eigenstates of the second class of $H_0$, that is, $N^{D}_{2}$ 
eigenstates of $H_{0}$. However, the mixing is small and local. 
That is, the Hamiltonian matrix of $H$ based on the $N^2_{D}$ eigenstates becomes a block matrix with many small blocks. 
This implies that the obtained eigenstates of the Hamiltonian matrix $H$ are low-entangled, where the deviation 
of the EE of one of the eigenstates of $H_{0}$ is small.

\section*{Appendix B. Numerical verification for identification of eigenstates and their LIOM-eigenvalues in a large system}
\begin{figure*}[t]
\begin{center} 
\includegraphics[width=14cm]{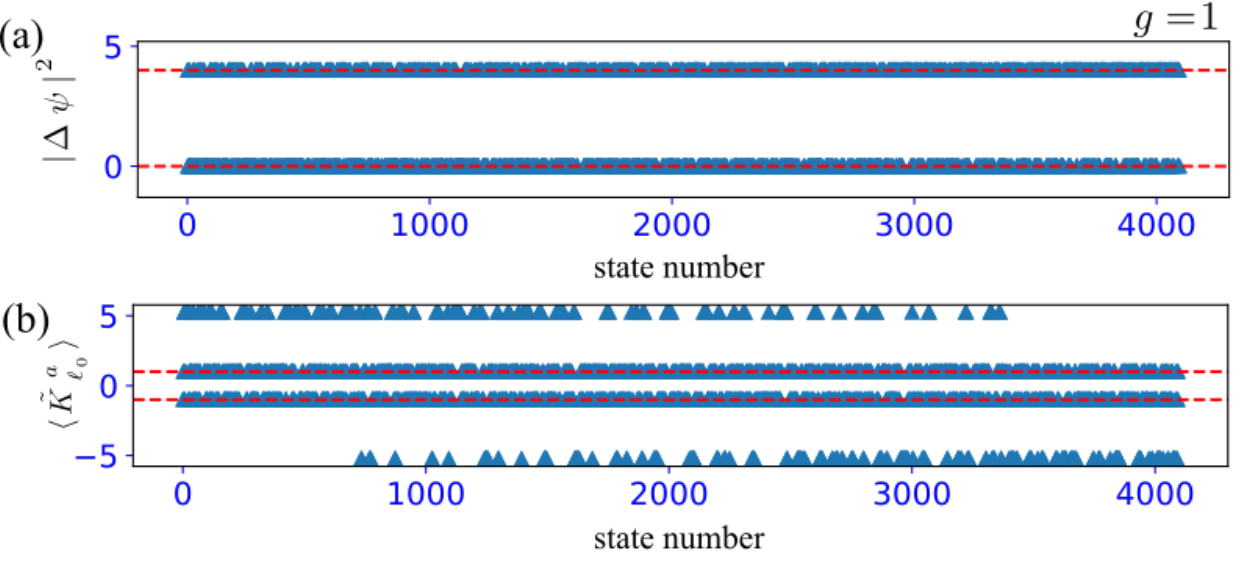}  
\end{center} 
\caption{Identification of energy eigenstates and their LIOM-eigenvalues for six unit-cell system ($L=12$). 
(a) $|\Delta \psi_k|^2$ for each eigenstate for $g=1$ case. 
(b) Each eigenvalue of $\tilde{K}^{a}_{\ell_0}$ for finite $g$ case. 
We set $\ell_0=1$.
All results are obtained for a single-shot disorder.
Energy eigenstates are numbered in the ascendant order.}
\label{Fig5}
\end{figure*}

In Sec.~IV, we showed how the modified LIOMs of Eq.~(\ref{Model_mod}) 
characterize eigenstates of the system for a small system size with analytical discussion.
In this appendix, we show a numerical verification of larger system size, $L=12$. 
The numerical results of $|\Delta \psi_{k}|^2$ and 
$\langle \psi_{k}|\tilde{K}^{a}_{\ell_0}|\psi_{k}\rangle$ are shown in 
Figs.~\ref{Fig5} (a) and ~\ref{Fig5} (b), where we set $g=1$ and $\ell_0=1$.
Even for large system size, all eigenstates are eigenstates for the modified LIOMs with some finite eigenvalues. 
The data imply that the modified LIOMs provide good quantum numbers for any system size $L$.

%%%%%%%%%%%%%%%%%%%%%%%%%%%%%%%%%%%% 
%%%%%%%%%%%%%%%%%%%%%%%%%%%%%%%%%%%%

\section*{Appendix C. Effect of fictitious disorder $\delta h_{\ell}$}

\begin{figure}[t]
\begin{center} 
\includegraphics[width=12cm]{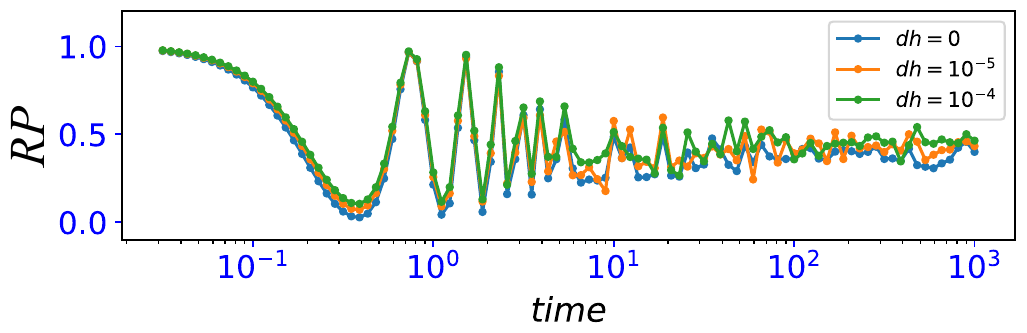}  
\end{center} 
\caption{Dependence on the strength of `fictitious' disorder $\delta h_{\ell}$ of 
the return probability with $\delta h_{\ell}\in[-\delta h,\delta h]$ (uniform distributed
disorder).
We set $L=12$, $W=2$, $g=1$ and $v_{zz}=0$. 
The initial state is a random cluster-basis state.
We averaged over 80 disorder and random initial state samples.}
\label{FigA5}
\end{figure}

To understand the structure of the eigenstate of the Hamiltonian $H$ of Eq.~(\ref{Model_mod}), we added a very small `fictitious' disorder $\delta h_{\ell}$.
We expect that such a small $\delta h_{\ell}$ gives little effect to the localization
nature of the system. 
As a concrete examination on this point, we observe the dependence on $\delta h$ 
of the return probability. 
The numerical estimation is shown in Fig.~\ref{FigA5}. 
The obtained dynamics of the return probability is almost independent of the strength 
of $\delta h$ where $\delta h \leq \mathcal{O}(10^{-4}g)$. 
From this fact, we expect that other physical observables in the system dynamics are 
not affected by the `fictitious' disorder $\delta h_{\ell}$.

%%%%%%%%%%%%%%%%%%%%%%%%%%%%%%%%%%%% 
%%%%%%%%%%%%%%%%%%%%%%%%%%%%%%%%%%%%

%%%%%%%%%%%%%%%%%%%%%%%%%%%%%%%%%%%% 
\begin{figure}[t]
\begin{center} 
\includegraphics[width=13.5cm]{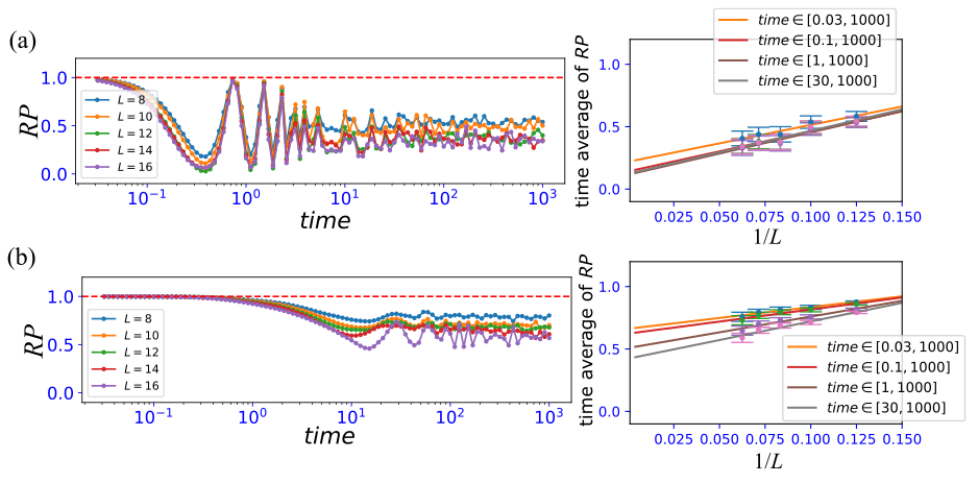}  
\end{center} 
\caption{System-size dependence of the return probability for $L=8,10,12$, $14$ and $16$. 
We set $W=2$, $\delta h=0$ and $g=1$. 
(a) The behaviors of the return probability for $v_{zz}=0$ case. 
We averaged over disorder and random cluster-basis state samples, the number of the 
disorder realization is 100, 80, 60, 40 and 20 for $L=8,10,12,14$ and $16$, respectively. 
Right panel: The system-size dependence of the time average of the return probability 
for the different time intervals, $t\in [0.03,10^{3}]$, $[0.1,10^{3}]$, $[1,10^{3}]$ and $[30,10^{3}]$. 
(b) The behaviors of the return probability for $v_{zz}=0.1$ case. We averaged over disorder where the initial state is the cluster-basis Neel state, the number of the disorder realization
is 100, 80, 60, 40 and 20 
for $L=8,10,12,14$ and $16$, respectively.
Right panel: The system-size dependence of the time average of the return probability 
for the different time intervals, $t\in [0.03,10^{3}]$, $[0.1,10^{3}]$, $[1,10^{3}]$ and $[30,10^{3}]$.}
\label{FigA6}
\end{figure}
%%%%%%%%%%%%%%%%%%%%%%%%%%%%%%%%%%%%

\section*{Appendix D. System-size dependence of return probability, loop order and modified loop order} 
In this appendix, we show the system-size dependence of the dynamics in detail. 
In particular, we focused on the return probability and calculated the return probability 
for various system sizes. 
The result without the Ising interaction ($v_{zz}=0$) is shown in Fig.~\ref{FigA6} (a). 
From the left panel of Fig.~\ref{FigA6} (a), 
the system-size dependence is small. 
All return probabilities remain at some finite value ($\sim 0.5$) for a long period. 
We expect that the finite value of the return probability also survives for larger 
system sizes as shown in the right panel of Fig.~\ref{FigA6} (a), where the dependence of the choice of the time interval in the time averaging operation is also small. 
We also observe the similar behavior even for a finite Ising interaction
($v_{zz}=0.1$), 
as shown in Fig.~\ref{FigA6} (b). 
The return probability for long times remains finite even for large system size. 
We expect that the finite value of the return probability also survives for larger 
system sizes as shown in the right panel of Fig.~\ref{FigA6} (b), 
where the dependence of the choice of the time interval in the time averaging operation is also small. 
These numerical results imply that the information of the initial state is preserved 
for long times even for large systems.

\begin{figure}[t]
\begin{center} 
\includegraphics[width=10cm]{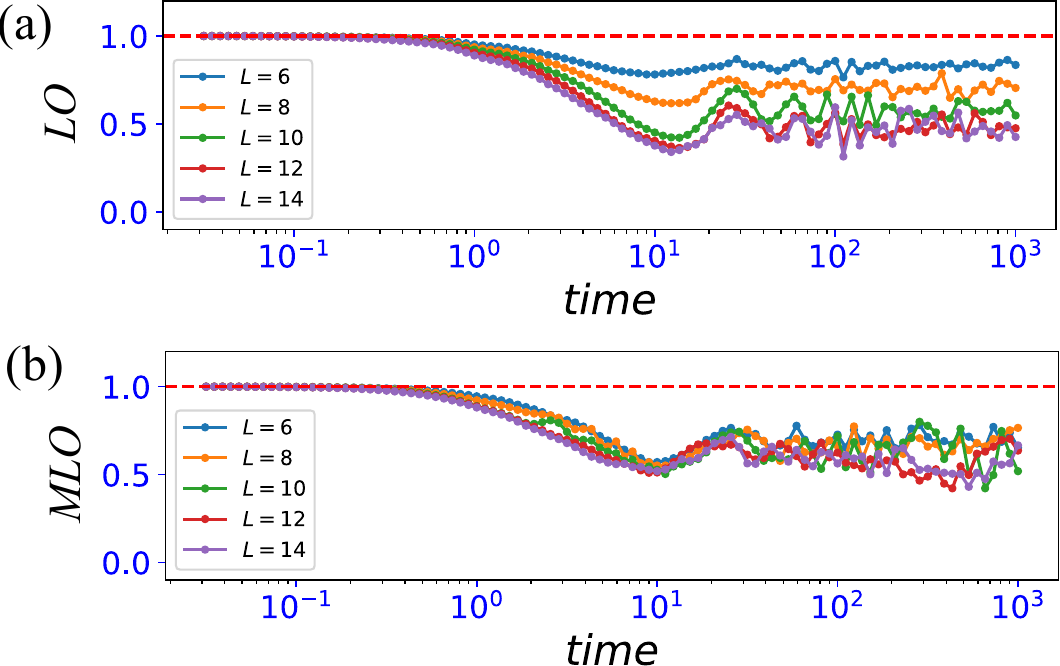}  
\end{center} 
\caption{System-size dependence of the loop order [(a)] and modified loop order [(b)] for $L=6,8,10$, $12$ and $14$.
We focus on $v_{zz}=0.1$ and set $W=2$, $\delta h=0$ and $g=1$. We averaged over disorder where the initial state is a cluster-basis Neel state and the number of the disorder realization is 100, 80, 60, 40 and 20 for $L=6,8,10,12$ and $14$, respectively.
We take the absolute value of the loop order and MLO for an odd $L/2$.}
\label{Fig9A}
\end{figure}

We further observed the system-size dependence of the time evolution of
the loop order for $v_{zz}=0.1$ as shown in Fig.~\ref{Fig9A} (a). 
Up to $t\sim 10^3$, the system-size dependence is small. The time average is shown in Fig.~\ref{Fig3_2} (c) in the main text indicates that the finite value of the loop order remains finite for larger systems. 

We also observe the system-size dependence of the dynamics of the MLO for $v_{zz}=0.1$ as shown in Fig.~\ref{Fig9A} (b). 
Up to $t\sim 10^3$, the system-size dependence is small. 
The time average as shown in Fig.~\ref{Fig3_2} (d) in the main text indicates that the finite value of the MLO remains for larger systems. 

%%%%%%%%%%%%%%%%%%%%%%%%%%%%%%%%%%%% 
%%%%%%%%%%%%%%%%%%%%%%%%%%%%%%%%%%%%

%%%%%%%%%%%%%%%%%%%%%%%%%%%%%%%%%%%%
%Fig
%\widetext
%\begin{figure}[t]
%\centering
%\includegraphics[width=14cm]{Fig10.pdf}
%\caption{System-size dependence of the modified loop order for
%$L=6,8,10$, $12$ and $14$.
%We fixed $v_{zz}=0.1$ and averaged over disorder where the %initial state is 
%the cluster Neel state, the number of the disorder %realization is 100, 80, 60, 40 and 20 
%for $L=6,8,10,12$ and $14$, respectively.
%Right panel: The system-size dependence of the time average of the modified loop orders 
%in from $t=0.03$ to $t=10^{3}$.
%}
%\label{FigA8}
%\end{figure}
%%%%%%%%%%%%%%%%%%%%%%%%%%%%%%%%%%%%

\section*{Appendix E. Level spacing analysis for finite $v_{zz}$}

\begin{figure}[t]
\centering
\includegraphics[width=10cm]{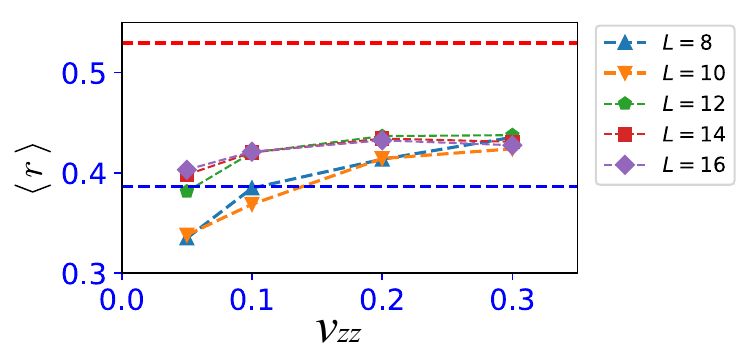}
\caption{
Mean level-spacing ratio $\langle r\rangle$ for the system with $L=8-16$, $W=2$, $g=1$ and $\delta h_{\ell}=0$. 
For small $v_{zz}$, $\langle r \rangle$ is close to the value of Poisson distribution. 
The number of the disorder realization is $100$, $60$, $40$,  $30$ and $10$ 
for $L=8$, $10$, $12$ $14$ and $16$, respectively. 
The red and blue dashed line are $\langle r\rangle=0.529$ and $0.386$, which are the values for 
the Wigner-Dyson and Poisson distributions, respectively.}
\label{FigA7}
\end{figure}

To examine the presence of the localization tendency and the integrability of the system, 
we employed the level spacing analysis for the system with finite $v_{zz}$ \cite{Oganesyan2007}.
We diagonalize the Hamiltonian $H+V_{zz}$, obtain all energy eigenvalues and calculate the level spacing ratio $r_s$ defined by $r_{s}=[{\rm min}(\delta^{(s)}, \delta^{(s+1)})]/[{\rm max}(\delta^{(s)},\delta^{(s+1)})]$ for all $s$, 
where $\delta^{(s)}=E_{s+1}-E_{s}$ and 
$\{E_{s}\}$ is the set of energy eigenvalue in ascending order. 
Then, we calculate the mean level spacing ratio $\langle r\rangle$, which is obtained by averaging over $r_{s}$ 
with employing all energy eigenvalues and also further averaging over disorder realizations for
$\{\lambda_{\ell}\}$. 
The result for various system sizes and $v_{zz}$ is shown in Fig.~\ref{FigA7}. 
For small $v_{zz}$, the remnant of degeneracy causes the mean value of the level spacing ratio to 
be smaller than that of the Poisson distribution, $2\ln 2 - 1\sim 0.386$. However, 
for larger $v_{zz}$, the value is getting slightly larger than the Poisson distribution but stays near
the value of the Poisson distribution or does not reach the value of the Wigner-Dyson distribution, $\sim 0.529$. 
This indicates that the system is in a localized phase.

%%%%%%%%%%%%%%%%%%%%%%%%%%%%%%%%%%%%
%Fig
%\widetext
%%%%%%%%%%%%%%%%%%%%%%%%%%%%%%%%%%%%

%\section*{\changeddd{Appendix F}. Calculation of the modified loop order}

%\changeddd{(3-27) In detail, this large oscillations come from the breakdown of the cluster-based Neel state (initial state) in the dynamics The initial cluster-based Neel state has the eigenvalue $+1$ even for each $\tilde{K}^{a}_{\ell}$ in the MLO. Thus, the initial value of the MLO is $+1$.   
%Along the time evolution, the initial state is deformed, then the second term of $\tilde{K}^{a}_{\ell}$ acts to the time evolved state, depending on the disorder value $\lambda_{\ell}$. 
%A large amplitude can appear in the case with a large $v_{zz}$ and a small $\lambda_{\ell}$. 
%The large amplitude induces a large value of the MLO as a whole.}
%We further observe the system-size dependence of the dynamics of 
%the MLO as shown in Fig.~\ref{FigA8} (b). 
%Up to $t\sim 10^3$, the system-size dependence is small. 
%The time average in the left panel of Fig.~\ref{FigA8} (b) indicates that the finite value of the MLO remains for larger systems. 
%Together with the results of Figs.~\ref{FigA6} (b) and \ref{Fig9A}, the presence of SPT 
%in the MBL is implied in the system with weak $V_{zz}$.

\section*{Appendix F. Novel LIOMs beyond the LIOMs $\tilde{K}^{a(b)}_\ell$}

As mentioned in Sec.VII, we can consider another type of LIOMs, 
which have not compact but a long-tail support.
They are introduced by changing 
$H_{\rm int}$ to $H'_{\rm int}=\sum_{r,\ell}V^{r}_{\ell}$, where
$V^{r}_{\ell}\equiv g e^{-|r|}N_{\ell-r-1}(K^{a+}_{\ell}K^{b-}_{\ell}+K^{a-}_{\ell}K^{b+}_{\ell})$. 
Then, the novel type of LIOMs can be constructed such as 
$L^{a}_{\ell}\equiv K^{a}_{\ell}+\frac{1}{2\lambda_{\ell}}\sum_{r}V^{r}_{\ell}$ 
and $L^{b}_{\ell}\equiv K^{b}_{\ell}-\frac{1}{2\lambda_{\ell}}\sum_{r}V^{r}_{\ell}$.
As a simplest case for the extension of $\tilde{K}^{a(b)}_{j}$, we include only $r=0$ 
and $1$ term. 

\begin{figure*}[h]
\begin{center} 
\includegraphics[width=14cm]{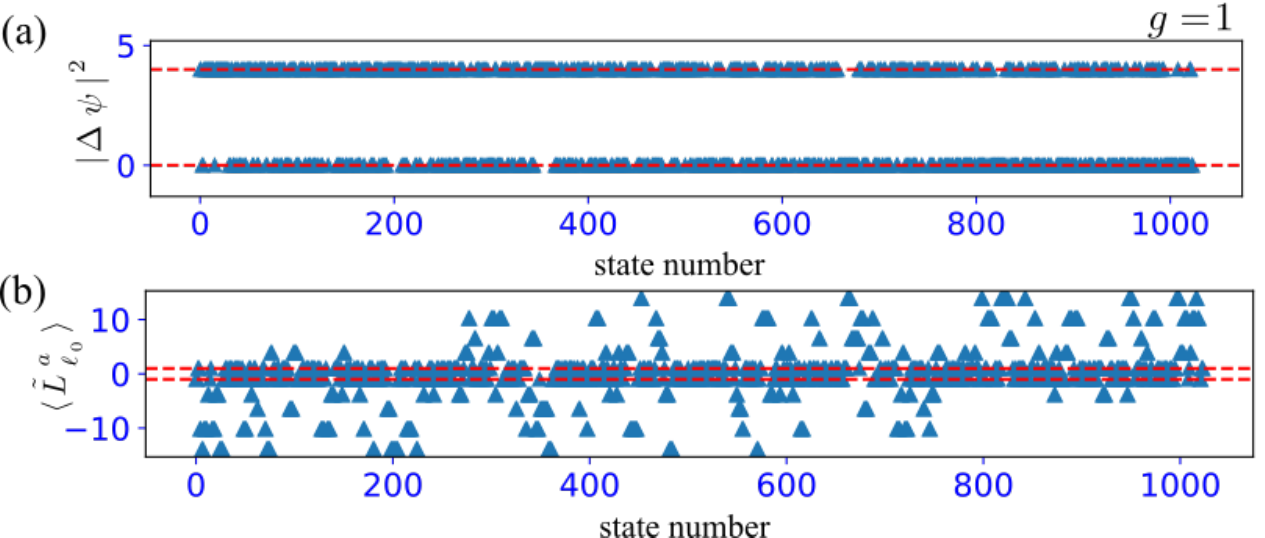}  
\end{center} 
\caption{Identification of eigenstates and $\tilde{L}^{a}_{\ell}$ LIOM-eigenvalues for five unit-cell system ($L=10$). 
(a) $|\Delta \psi_k|^2$ for each eigenstate for $g=1$ case. 
(b) Each eigenvalue of $\tilde{L}^{a}_{\ell_0}$ for $g=1$ case. We set $\ell_0=1$. 
These results are obtained for a single-shot disorder.
Energy eigenstates are numbered in the ascendant order.}
\label{Fig11}
\end{figure*}

The novel LIOMs are given as 
\begin{eqnarray}
\tilde{L}^a_{\ell}=K^{a}_{\ell}+\frac{g}{2\lambda_{\ell}}
[V^{r=0}_{\ell}+V^{r=1}_{\ell}],\:\:\: 
\tilde{L}^b_{\ell}=K^{b}_{\ell}-\frac{g}{2\lambda_{\ell}}
[V^{r=0}_{\ell}+V^{r=1}_{\ell}]\nonumber.
\label{LIOM3}
\end{eqnarray}
Whether the above operators actually play a role of the LIOMs 
can be examined numerically in the same way as the numerical calculations 
in Fig.~\ref{Fig1} (e) and ~\ref{Fig1} (f). 
In numerical calculation, we set $L=10$. 
The numerical results of $|\Delta \psi_{k}|^2$ and 
$\langle \psi_{k}|\tilde{L}^{a}_{\ell_0}|\psi_{k}\rangle$ are shown in Fig.~\ref{Fig11} 
(a) and ~\ref{Fig11} (b), where we set $g=1$ and $\ell_0=1$. 
We observed that certainly all energy eigenstates are eigenstates for the LIOMs $\tilde{L}^{a(b)}_{\ell}$ with some finite eigenvalues.

%%%%%%%%%%%%%%%%%%%%%%%%%%%%%%%%%%%% 
%%%%%%%%%%%%%%%%%%%%%%%%%%%%%%%%%%%%


\begin{thebibliography}{10}
\bibitem{clusterEE}
This value of the EE corresponds to the value when the two cluster states are cut.
\end{thebibliography}
\end{document}